\pdfoutput=1

\documentclass[11pt]{article}

\usepackage{acl}

\usepackage{times}
\usepackage{latexsym}

\usepackage[T1]{fontenc}
\usepackage[utf8]{inputenc}

\usepackage{microtype}

\usepackage{amsmath}
\usepackage{amssymb}
\usepackage{blindtext}

\usepackage{booktabs}
\usepackage{subcaption}
\usepackage{graphicx}
\usepackage[frozencache=true,cachedir=minted-cache]{minted} 
\usepackage{multirow}
\usepackage[english, ruled,vlined, boxed]{algorithm2e}
\usepackage{algpseudocode}
\usepackage{dirtytalk}

\title{Addressing Leakage in Self-Supervised Contextualized Code Retrieval}

\author{Johannes Villmow  \and Viola Campos  \and Adrian Ulges \and Ulrich Schwanecke \\
  RheinMain University of Applied Sciences \\ Wiesbaden, Germany \\ \texttt{\{firstname.lastname\}@hs-rm.de}}

\begin{document}
\maketitle

\begin{abstract}
We address contextualized code retrieval, the search for code snippets helpful to fill gaps in a partial input program. Our approach facilitates a large-scale self-supervised contrastive training by splitting source code randomly into contexts and targets. To combat leakage between the two, we suggest a novel approach based on mutual identifier masking, dedentation, and the selection of syntax-aligned targets.  Our second contribution is 
a new dataset for direct evaluation of contextualized code retrieval, based on a dataset of manually aligned subpassages of code clones. Our experiments demonstrate that our approach improves retrieval substantially, and yields new state-of-the-art results for code clone and defect detection.
\end{abstract}

\section{Introduction}
AI-supported software development has experienced growing interest recently~\cite{lu21CodeXGLUE}. Its most intensely researched tasks include code auto-completion~\cite{svyatkovskiy2020codeGPT}, natural language code search~\cite{husain2019codesearchnet}, and code clone detection~\cite{svajlenko25bigclonebench}. All these tasks require a semantic understanding of code, and are commonly tackled by pretrained 
transformers~\cite{vaswani2017attention}.

Our focus is on a related task called {\it contextualized code search}~\cite{mukherjee20contextualized,dahal2022scotch}: Given an incomplete piece of code and a certain position of interest (e.g., the current cursor position), a retriever searches for 
 code fragments that are relevant for filling in the missing piece. 
 This setting aligns well with programmers' workflow
 , and differs substantially from the three above tasks: (1) In contrast to natural language code search, retrievers in contextualized code search can exploit local code context, which may (but does not have to) include a comment describing the missing step. 
 (2) Code generated by autocompletion
  (e.g. from GitHub's CodEx \cite{chen2021codex}) is prone to subtle programming errors that may alter the behaviour of the current program, particularly if the current code status still contains bugs or larger gaps. In contrast, contextualized search leaves 
  the developer in charge, 
and the origin of a solution remains transparent. 
 (3) In contrast to clone detection, contextualized code search is not targeted at semantically similar pieces of code but pieces of code that complement each other.

A key challenge with contextualized code search is that supervised labels for relevant code passages are missing. 
Therefore, we bootstrap a \emph{self-supervised learning process}  by drawing inspiration from Cloze Tasks in natural language processing~\cite{lee19inversecloze}: 
Given a large-scale dataset containing a piece of code, we erase a random block. We refer to this block as the {\it target}, and to the rest as {\it context}. 
Together, they both form a positive sample for contrastive learning.

\begin{figure*}[!ht]
\centering
\includegraphics[width=1\textwidth]{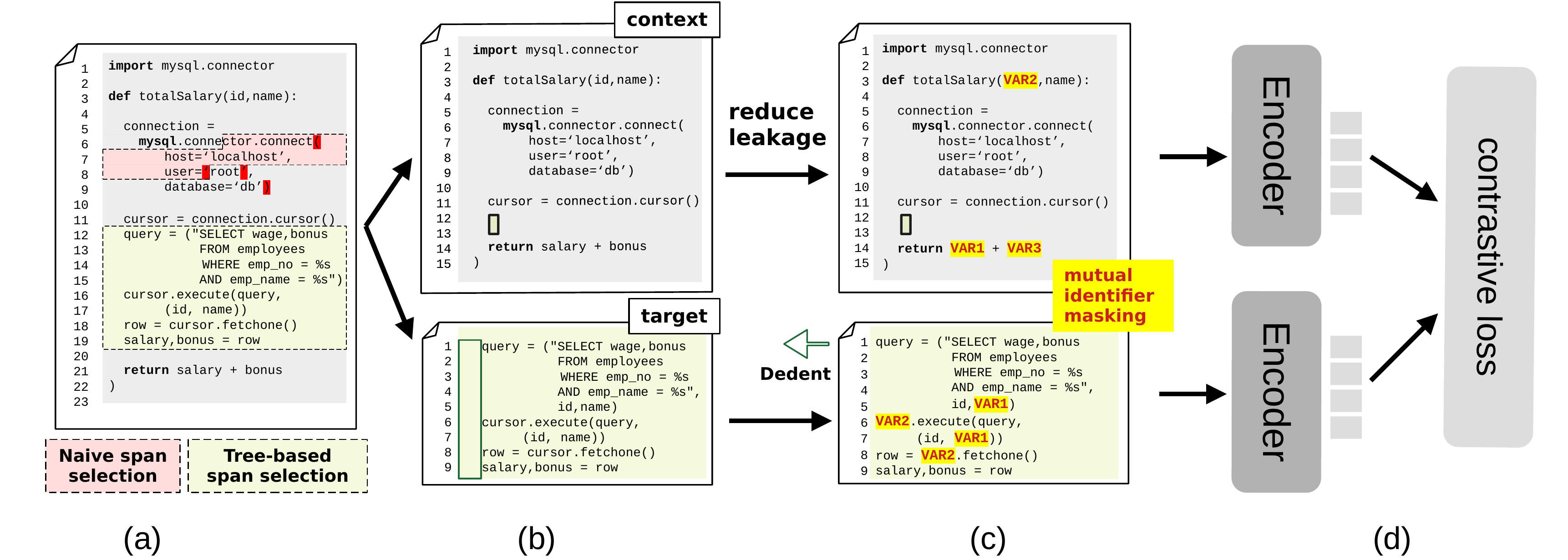}
\caption{Our approach bootstraps code pairs for contrastive learning from random source code blocks. To avoid leakage, such as cutting syntactic primitives (red), we (a) select the target using the code's syntactic structure (green). We split (b) the code into context (gray) and target (green). We further (c) mutually mask identifiers (yellow) and dedent the target (green arrow). Finally, we apply contrastive learning on the resulting code pairs (d).}
\label{fig:approach}
\end{figure*}

Unfortunately, this approach suffers from leakage between context and target (see Figure \ref{fig:approach}): 
(1) Both share common identifiers, 
(2) the target's indentation level matches the position of interest in the query, and 
(3) if context and target divide a syntactic primitive (e.g., a for-loop), the target can easily be identified by bracket matching the ones in the context. Retrievers might exploit all these effects and bypass semantic similarity.
To this end, our first contribution is a novel approach towards self-supervised code retrieval, which avoids the above bias through de-leaking steps, namely mutual identifier masking, dedentation, and target selection aligned with the code's syntactic primitives.

The second challenge we address is evaluation: So far, the focus of evaluating code retrieval systems has been on natural language queries (which can be bootstrapped from docstrings)~\cite{husain2019codesearchnet}. Contextualized code retrieval
has been evaluated only indirectly via infilling quality~\cite{lu2022reacc}, which poorly reflects the actual retrieval quality.
Therefore, our second contribution is a 
rigorous evaluation of contextualized code retrieval on a manually curated dataset based on aligned code clones. We coin \texttt{COCOS} and make available for future research.
We demonstrate on said dataset that retrieval quality benefits substantially from our de-leaking approach. Also, we achieve state-of-the-art results on the related tasks code clone and defect detection. 

\section{Approach}
\label{sec:approach}

Our system takes arbitrary pieces of code and bootstraps code pairs for contrastive learning. 
More specifically, given a piece of code as a token sequence 
$X{=}x_1,\dots ,x_n$,
we select target code snippet 
$Y{=}x_i,\dots ,x_{i+L}$ with $n,i,L \in \mathbb{N}$
, obtaining a masked context version 
$X'{=}x_1,\dots ,x_{i-1},x_{\mathrm{MASK}},x_{i+L+1},\dots,x_n$. To $X', Y$ we
prepend a language-specific CLS token.

To address the above mentioned leakages we utilize the concrete syntax tree of a piece of code\footnote{We use the tree-sitter library for parsing.}. The leaves of this tree consist of all code tokens including whitespace tokens.
First we propose \textbf{tree-based span selection} (TS) 
and use a program's syntax tree to determine the cut-out target span $Y$ (see Figure \ref{fig:approach}). 
By selecting syntax tree nodes for removal, we ensure to always remove a syntactically complete piece, using the following procedure: 
We sample the target's length $L$ from a normal distribution with $\mu{=}150$ and $\sigma{=}90$. Then select a node $n$ in the syntax tree  covering at most $L$ leaves/tokens and iteratively expand the selection, either to $n$'s parent, or by adding $n$'s direct siblings, until reaching the desired size $L$.
Adding siblings allows for multiline targets spanning several statements (but not a complete block).
Second, we suggest 
\textbf{ mutual identifier masking (IM)}\footnote{What is an identifier is defined in the grammar of a tree-sitter parser and varies between programming languages. I.e. we do not differentiate between variables, method names or method calls.}:
By modifying the syntax tree, we replace identifiers such as variable names with special tokens (e.g. VAR1, VAR2) 
dynamically during training,
resulting in a more structure-oriented approach. 
To minimize not only masking but at the same time leakage we mask only mutual identifiers present in both context and target. We hide 90\% of those mutual identifiers randomly either in the context or in target code. For 5\% of context-target pairs we omit identifier masking overall. 

Finally, we determine the indentation level of the target code and \textbf{dedent} (DE) it, so that it has indentation level zero. 

\subsection{Training}

We encode both sequences with the same 
transformer encoder and obtain sequence embeddings $\mathbf{q}, \mathbf{k} \in \mathbb{R}^d$ for context code $X'$ and target $Y$
by using the encoding of the CLS token. Following \citet{wang-etal-2021-codet5} we optionally pretrain the transformer with alternating generation tasks such as identifier masking and span prediction\footnote{We omit identifier detection and instead use our tree-based span selection to generate large and small spans.} .
The retriever is then trained by optimizing the contrastive InfoNCE \cite{oord2018infoNCE} 
loss with in-batch negative samples :
\begin{equation}
\label{eq:infonce_loss}
    \mathcal{L}_\Theta = - log 
    \frac{
       \text{exp}(f(\mathbf{q}, \mathbf{k}^+) / \tau ))
    }
    {
        \sum_{i=0}^{K-2} \text{exp}(f(\mathbf{q}, \mathbf{k}^-_i) / \tau ))
    }
\end{equation}
Where $f$ is cosine similarity, $K$ is the amount of sequences in our batch and $\tau{=}0.1$ temperature. 
To obtain harder negative samples - which have been found crucial for good retriever training \cite{ren-etal-2021-rocketqav2} - we form batches only with samples from the same programming language.

\section{Dataset}

Our self-supervised code retrieval model is pre-trained on 33M files in 16 programming languages (see Appendix \ref{sec:pretraining_dataset}). 
As code files tend to be large, we truncate them using \textbf{tree-based span selection} (compare Section \ref{sec:approach}): Starting from a file we randomly select sufficiently large spans of code 
(length between 150 and 800 tokens). 
We remove those segments from the original file, and feed 
the shortened file as well as all individual segments as inputs $X$ into the learning process described in Section \ref{sec:approach}. A special identifier (similar to code folding in the IDE) marks those positions in the original file where segments have been removed.

\subsection{Contextualized Code Retrieval Dataset}

 Evaluating contextualized code retrieval models is hard, due to limited to no suitable evaluation data being available indicating which subblocks in code implement the same functionality. To address this gap, we have created a new dataset based on BigCloneBench~\cite{svajlenko25bigclonebench} -- a code clone dataset that provides pairs of functions implementing the same functionality\footnote{In that dataset two functions are considered clones, but the shared functionality may only be a part of the function. We use the shared functionality part as relevant targets and obtain different contexts.}. 
 We manually select a sub-passage from a function as the target and label which lines in the function's clones match this target (see Figure \ref{fig:cocos_context} - \ref{fig:cocos_solution}).
 By extracting these targets and their surrounding contexts, we can evaluate how well a model retrieves targets implementing the same functionality. Note that the original target is not considered. We manually gather $387$ context-target pairs implementing 20 randomly selected functionalities and coin the resulting dataset \texttt{COCOS} ( {\bf Co}ntextualized {\bf Co}de {\bf S}earch).

\begin{table}[htb]
    
        \centering
        \resizebox{\linewidth}{!}{%
        \begin{tabular}{@{}lrrrrr@{}}
        \toprule
        \textbf{Model Features} & \multicolumn{1}{l}{\textbf{MAP}} & \multicolumn{1}{l}{\textbf{NDCG}} & \multicolumn{1}{l}{\textbf{P@1}} & \multicolumn{1}{l}{\textbf{P@3}} & \multicolumn{1}{l}{\textbf{P@10}} \\ \midrule
        TS, PT & 50.04 & 77.71 & \textbf{78.12} & 69.97 & 56.56 \\
        TS, IM, DE & 56.07 & 76.24 & 38.8 & 42.8 & 54.58 \\
        TS, IM, DE, PT & \textbf{69.14} & \textbf{85.85} & 74.48 & \textbf{77.95} & \textbf{73.72} \\ \bottomrule\end{tabular}%
        }
        \caption{Zeroshot code retrieval results for different deleaking steps: Tree-based span selection (TS); mutual identifier masking (IM); dedenting (DE); pretraining (PT). Compare Section \ref{sec:approach}.}
        \label{tab:results-zeroshot}
\end{table}

\section{Evaluation}

We evaluate our model on zero-shot code retrieval on our \texttt{COCOS} dataset and on two similar code understanding tasks from CodeXGlue ~\cite{lu21CodeXGLUE}, specifically code clone detection and code defect detection. 
During unsupervised training we measure mean reciprocal rank (MRR) on $30k$ samples of the held out validation set (for which we apply the same modifications). For all experiments we report test results of the model with the highest validation MRR. 

\subsection{Zeroshot Code Retrieval}

Without further fine-tuning we directly evaluate our self-supervised model on the \texttt{COCOS} dataset in a zero-shot setting. For each context we rank all possible targets, exclude the original target. In Table \ref{tab:results-zeroshot} we report mean average precision (MAP), normalized discounted cumulative gain (NDCG) and precision at $k$. The baseline trained without deleaking steps dedentation and mutual identifier masking is able to retrieve samples with similar identifiers (high precision at 1), but fails to consistently retrieve all relevant targets. Instead our approach -- that uses deleaking steps -- retrieves more relevant samples. This is visible in Figure \ref{fig:tsne}, where our approach  forms better clusters for both contexts and targets.

\subsection{Clone Detection}
We evaluate our model on clone detection on the \texttt{POJ-104} dataset \cite{mou2016convolutional}. The \texttt{POJ-104} dataset consists of C and C++ programs for 104 problems from an open programming platform (OJ). 
We follow the evaluation procedure of CodeXGlue and report mean average precision (MAP@R) with R=499. We find that our model outperforms state-of-the-art by a large margin. 

\begin{table}[htb]

        \centering
            \begin{tabular}{@{}lrr@{}}
            \toprule
            \multicolumn{1}{c}{\multirow{2}{*}{\textbf{Model}}} & \multicolumn{1}{c}{\textbf{Clone}} & \multicolumn{1}{c}{\textbf{Defect}} \\ \cmidrule(l){2-2} \cmidrule(l){3-3} 
            \multicolumn{1}{c}{} & \multicolumn{1}{c}{\textbf{MAP@R}} & \multicolumn{1}{c}{\textbf{Accuracy}} \\ \midrule
            RoBERTa (code) & 76.67 & 61.05 \\
            CodeBERT & 82.67 & 62.08 \\
            code2vec & 1.98 & 62.48 \\
            PLBART & - & 63.18 \\
            GraphCodeBERT & 85.16 & 63.21 \\
            SynCoBERT & 88.24 & 64.50 \\
            C-BERT & - & 65.45 \\
            CodeT5 & - & 65.78 \\
            CoTexT & - & 66.62 \\ \midrule
            Ours & \textbf{91.34} & ~\textbf{68.33} \\ \bottomrule
            \end{tabular}%
        \caption{Results on code clone detection on \texttt{POJ-104} and defect detection on \texttt{Devign} dataset. }
        \label{tab:results-clone-detection}
\end{table}
\subsection{Defect Detection}
We evaluate our model on defect detection on the \texttt{Devign} dataset \cite{zhou-2019-devign}. It consists of vulnerable C functions manually collected from open source projects. The task is to predict whether the function is vulnerable. Following CodeXGlue we report accuracy.

\section{Related Work}

Given the success of pre-trained language models in NLP, some recent work has extended pre-training to program syntax. \citet{kanade2020cubert} and \citet{feng2020codebert} train a BERT 
 encoder on source code using masked language modeling. \citet{Guo2021GraphCodeBERT} propose GraphCodeBERT to incorporate structural information like data flow. 
Besides these encoder models, \citet{svyatkovskiy2020codeGPT} and \citet{liu2020cugLM} developed CodeGPT and CugLM, both based on the transformer decoder and pre-trained on pairs of natural language and program code. \citet{ahmad-etal-2021-unified} and \citet{wang-etal-2021-codet5} proposed PLBART and CodeT5, following the architecture of BART 
and T5 respectively, trained on functions in multiple programming languages paired with natural language comments. SynCoBERT \cite{wang2021syncobert} is trained on various pre-training tasks on multi-modal data, including code, comment and AST representations. 
\citet{guo2022unixcoder} propose UniXcoder, which takes a similar approach but employs a encoder-decoder architecture instead of a single encoder.   

In most of the above work, multiple modalities have been applied, e.g. code and natural language comments. 
In contrast to contextual code search, this setup does not come with leakage, which is the main concern of our work.

\paragraph{Code-to-code search} A recent line of work uses program code as additional context for natural language queries to clarify the programmer's intent. FaCoY \cite{kim2018facoy} extends the query with related code from StackOverFlow and searches similar code fragments in the source code index. 
SCOTCH \cite{dahal2022scotch} studies query augmentation with surrounding source code to improve search results. \citet{jain2021} propose ContraCode, which introduces compiler-based semantic-preserving code transformations as data augmentations and trains the neural network on a contrastive learning objective comparing similar and dissimilar code snippets.
Plain code-to-code search is investigated in \cite{ragkhitwetsagul2019siamese}, where code fragments are used as query to find similar code clones in a code base. To improve performance, multiple code representations are combined to capture code structure at different levels.
Aroma \cite{luan2019aroma} clusters candidate code and intersects the snippets in each cluster to recommend likely subsequent code for a given snippet.

\paragraph{Contextualized Code Search}
\citet{muk2020codec} address contextualized code search 
by decompiling code fragments into a simpler representation called SKETCH \cite{murali2017bayesian} to learn a statistical search model. 
\citet{lu2022reacc} propose ReACC and use partial code as search query in the context of retrieval-augmented code completion. They focus on improving generation with results from a contrastive retriever. Differing to ours they do not train to retrieve partial targets, but aim to retrieve the full augmented file inserting dead code and renaming variables to combat leakage. However, even with dead code insertion a strong structural overlap between query and target exists. Compared to ContraCode and ReACC our steps towards leakage reduction are much simpler. 


\section{Conclusion}

We have proposed a new approach towards unsupervised code retrieval, which reduces leakage between randomly drawn targets and their contexts.
We also contribute a dataset \texttt{COCOS}, 
on which we demonstrate via ablations that leakage reduction is crucial for an efficient training. Also, our approach yields competitive representations for related tasks, as demonstrated by new state-of-the-art results on clone and defect detection.
An interesting future direction will be to combine our retriever with generators for a combined, unsupervised trainig.

\bibliographystyle{acl_natbib}

\appendix

\section{Pre-training Dataset Details}
\label{sec:pretraining_dataset}

We crawl 237k active GitHub repositories with more than 10 stars\footnote{We consider a repository as active if there has been a pull request between 04/21 and 09/21 } and perform per file deduplication. We keep files in programming languages for which a tree-sitter parser is available (16 languages). The resulting dataset is shown in Table \ref{tab:dataset} and consists of $\approx33M$ code files in 16 programming languages. We select 570 repositories for validation.

\begin{table}[htb]
\centering
\begin{tabular}{@{}lrrr@{}}
\toprule
\multicolumn{1}{c}{\textbf{Language}} & \multicolumn{1}{c}{\textbf{Training}} & \multicolumn{1}{c}{\textbf{Valid}} & \multicolumn{1}{c}{\textbf{Total}} \\ \midrule
Java                                  & 7,345,753                             & 8,434                              & 7,354,187                          \\
JavaScript                            & 4,471,689                             & 14,134                             & 4,485,823                          \\
C++                                   & 3,734,357                             & 1,698                              & 3,736,055                          \\
Python                                & 3,016,545                             & 4,718                              & 3,021,263                          \\
C\#                                   & 2,843,642                             & 570                                & 2,844,212                          \\
TypeScript                            & 2,299,964                             & 2,392                              & 2,302,356                          \\
C                                     & 2,242,379                             & 781                                & 2,243,160                          \\
PHP                                   & 2,206,063                             & 4,648                              & 2,210,711                          \\
Go                                    & 1,759,600                             & 129                                & 1,759,729                          \\
Ruby                                  & 1,068,668                             & 3,397                              & 1,072,065                          \\
Rust                                  & 366,891                               & 54                                 & 366,945                            \\
CSS                                   & 349,525                               & 2,579                              & 352,104                            \\
Scala                                 & 273,822                               & 1,198                              & 275,020                            \\
Haskell                               & 114,311                               & 177                                & 114,488                            \\
OCaml                                 & 55,838                                & 0                                  & 55,838                             \\
Julia                                 & 34,403                                & 29                                 & 34,432                             \\ \bottomrule
\end{tabular}
\caption{Number of files in unsupervised pre-training dataset.}
\label{tab:dataset}
\end{table}

\section{Training Details}

On all models and tasks we use the AdamW optimizer and linearly increase the learning rate for 10\% of the training steps, along with a polynomial decay for the remaining steps. 

We train our unsupervised models for 500k steps on a single A6000 GPU, with a peak learning rate of $0.0001$ and use a dynamic batch size so that batches contain around 7000 tokens. 

For clone and defect detection we fine-tune our model on the respective training set. Following \citet{wang-etal-2021-codet5} we run a brief sweep over learning rate, batch size and number of epochs and report results of the model with highest validation score, using the published evaluation code.

We will release our code in future.

\begin{figure*}[!ht]
\begin{minipage}[b]{.49\textwidth}
    \centering
    \includegraphics[width=\textwidth]{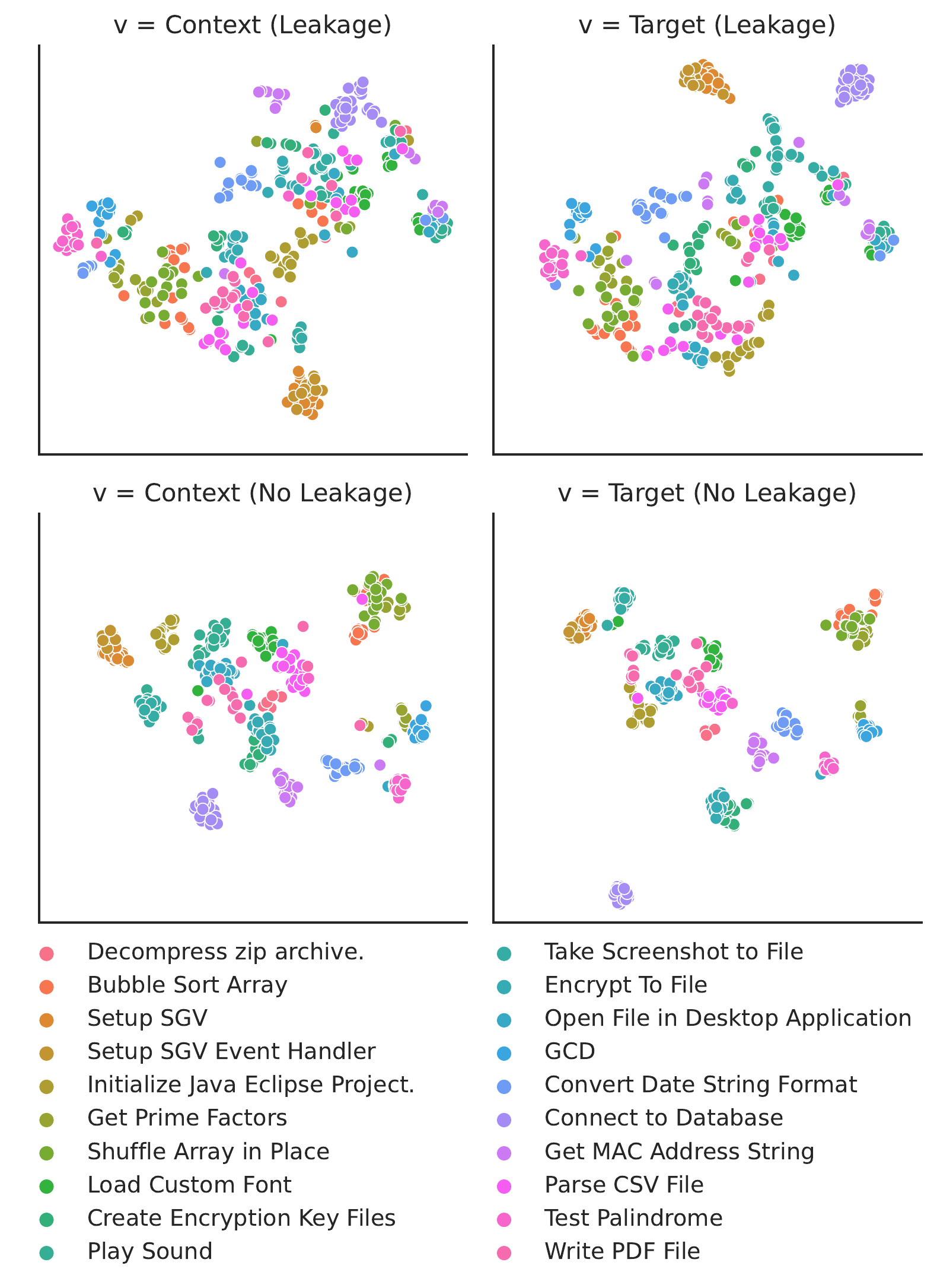}
    \caption{TSNE comparison between the embeddings of the baseline model with leakage (top) and our steps for leakage reduction have been applied (bottom).}
    \vspace{4ex}
    \label{fig:tsne}
\end{minipage} 
\hfill
 \begin{minipage}[b]{.49\textwidth}
 \centering
  \begin{minted}[
  frame=lines,
% framesep=2mm,
% baselinestretch=1.2,
fontsize=\tiny
]{java}
public boolean extract(File f, String folder) {
    Enumeration entries;
    ZipFile zipFile;
    try {
        zipFile = new ZipFile(f);
        entries = zipFile.getEntries();
        [MASK]
        zipFile.close();
    } catch (IOException ioe) {
        this.errMsg = ioe.getMessage();
        Malgn.errorLog(
            "{Zip.unzip} " + ioe.getMessage()
        );
        return false;
    }
    return true;
}
\end{minted}
\caption{Incomplete and masked query $X'$ from our \texttt{COCOS} dataset. The \texttt{[MASK]} token denotes the current position of interest (cursor). Code that extracts elements from a zip file needs to be found. It can be seen that our approach forms better clusters.}
\vspace{4ex}

\label{fig:cocos_context}
\end{minipage}%
\hfill
\begin{minipage}[b]{.49\textwidth}
\centering
\begin{minted}[
frame=lines,
framesep=2mm,
baselinestretch=1,
% bgcolor=monokai,
fontsize=\tiny
]{java}
while (entries.hasMoreElements()) {
    ZipArchiveEntry entry = 
        (ZipArchiveEntry) entries.nextElement();
    if (entry == null) continue;
    String path = folder + "/"
        + entry.getName().replace('\\', '/');
    if (!entry.isDirectory()) {
        File destFile = new File(path);
        String parent = destFile.getParent();
        if (parent != null) {
            File parentFile = new File(parent);
            if (!parentFile.exists()) {
                parentFile.mkdirs();
            }
        }
        copyInputStream(
            zipFile.getInputStream(entry), 
            new BufferedOutputStream(
                new FileOutputStream(destFile)
            )
        );
    }
}
\end{minted}
\caption{The masked section $Y$ manually selected from $X$ (Figure \ref{fig:cocos_context}). It has been dedented for better readability.}
\label{fig:cocos_target}
\end{minipage}\hfill
\begin{minipage}[b]{.49\textwidth}
\centering

\begin{minted}[
frame=lines,
% framesep=2mm,
% baselinestretch=1,
% bgcolor=monokai,
fontsize=\tiny
]{java}
ArchiveEntry ae = zis.getNextEntry(); 
while(ae != null) {
    //Resolve new file
    File newFile = new File(
        outputdir + File.separator + ae.getName()
    );
    
    //Create parent directories if not exists
    if(!newFile.getParentFile().exists())
        newFile.getParentFile().mkdirs();
    
    if(ae.isDirectory()) { //create if not exists
        if(!newFile.exists())
            newFile.mkdir();
    } else { //If file, write file
        FileOutputStream fos = new FileOutputStream(
            newFile);
        int len;
        while((len = zis.read(buffer)) > 0) {
            fos.write(buffer, 0, len);
        }
        fos.close();
    }
    
    //Proceed to the next entry in the zip file
    ae = zis.getNextEntry();
}
\end{minted}
\caption{Possible solution that implements the same functionality as the target in Figure \ref{fig:cocos_target}}
\label{fig:cocos_solution}
\end{minipage}%
\hfill

\end{figure*}

\begin{figure*}[htb]
    
\end{figure*}

\end{document}